\documentclass[showpacs,preprintnumbers,amsmath,amssymb]{revtex4}

\usepackage{graphicx}
\usepackage{dcolumn}
\usepackage{bm}

\begin{document}

\title{On the indistinguishability of Chiral QED with parameter-free Faddeevian anomaly and QED under a chiral constraint}

\author{Anisur Rahaman}
\affiliation{Durgapur Government College, Durgapur - 713214,
Burdwan, West Bengal, India}

\email{manisurn@gmail.com}

\date{\today}

\begin{abstract}
We carry out an investigation imposing a chiral constraint in the
phase space of vector and axial-vector Schwinger model. We find
that resulting model is identical to gauge non-invariant model
which was obtained by the imposition of chiral constraint in the
phase-space of in Chiral Schwinger model with the parameter-free
Faddeevian anomaly. Three different models having different types
of interaction between the matter and gauge field become
indistinguishable under a chiral constraint at the quantum
mechanical level. The resulting gauge non-invariant model has an
equivalent gauge invariant version in the same phase space that
can be identified with the vector Schwinger model.

\end{abstract}

\pacs{11.10.Ef, 11.30.Rd} \maketitle

\section{Introduction}
Imposition of chiral constraint has a remarkable feature in the
$(1 +1)$ dimensional field theoretical model. It was first
introduced in the seminal work of Harada \cite{KH}. Imposition of
chiral constraint puts a restriction on the degrees of freedom of
a boson with a particular chirality depending on the nature of
constraint. In the article \cite{KH}, it was shown that imposing a
chiral constraint in the phase-space of the chiral Schwinger model
\cite{JR, ROT1, ROT2} it could be expressed in terms of a chiral
boson, which was the basic ingredient of heterotic String theory
\cite{SIG, SONI}. The idea of the imposition of chiral constraint
has been used in different situations \cite{MIAO, ARAN, ARN, ARS,
AR1}. The fascinating role of imposition of chiral constraint to
save the phenomena of s-wave scattering off dilaton black hole
from the danger of information loss is observed in \cite{ARN}. In
the article \cite{AR}, we found that the model generated in
\cite{KH} by the imposition of the chiral constraint had its
origin in the gauge model of chiral boson \cite{BEL}. Apart from
the standard Lorentz co-variant one-parameter class of
regularization \cite{JR, RABIN}, in the article \cite{PM, MG} the
authors showed that the chiral Schwinger model was also found to
be physically sensible for a Lorentz non-covariant parameter-free
regularization which resulted in a Faddeevian type of anomaly
\cite{FAD1, FAD2}. The imposition of chiral constraint in the
phasespace of the theory enables us to express this model in terms
of chiral boson in \cite{ARN}. The resulting model in this
situation also finds its origin in the gauged version of the
chiral boson \cite{BEL}, when the co-variant mass masslike term
for the gauge field is replaced by a non-covariant masslike term
\cite{AR}. The role of the chiral constraint that has been studied
so far was restricted on the model where interaction is of chiral
nature. A natural question may arise whether the imposition of
chiral constraint on the models where the interaction is of vector
and axial-vector nature may lead to a physically sensible
field-theoretic model as it was found in the case of the model
when the interaction was of chiral nature. Keeping it in view, an
attempt has been made to impose, the chiral constraint on vector
and axial-vector Schwinger model \cite{SCH, LOW} and observe that
chiral Schwinger model with Faddeevian anomaly due to Mitra, the
vector Schwinger model, and axial-vector Schwinger model all map
onto a single gauge non-invariant model which we are going to
describe here. This gauge non-invariant model indeed has an
equivalent gauge-invariant version in the usual phase-space which
can be identified with the vector Schwinger model.

The article is organized as follows. In Sec.II, we have given a
brief review of the models in which we are going to impose the
chiral constraint. Sec. III is devoted to the imposition of chiral
constraint in the vector Schwinger model, axial-vector Schwinger
model, and the chiral Schwinger model with parameter-free
Fadeevian anomaly. In Sec.IV, we carry out the gauge-invariant
reformulation of the model, which has resulted after the
imposition of chiral constraint on the said models.  Sec. V has
devoted to the description of the theoretical spectra of the model
resulted after the imposition of the chiral constraint. Sec. VI
contains a brief discussion and conclusions.

\section{Brief review  of the models}
\subsection{Chiral Schwinger model}
The chiral Schwinger model is described by the fermionic
Lagrangian density
\begin{eqnarray}
{\cal L}_{CS} &=& \bar\psi\gamma^\mu[i\partial_\mu +
gA_\mu(1-\gamma_5)]\psi-\frac{1}{4} F_{\mu\nu}F^{\mu\nu}.
\label{FCS}
\end{eqnarray}
Here$\psi$ and $A_\mu$ are fermion and gauge fields respectively.
The field strength tensor is defined by $F_{\mu\nu} = \partial_\mu
A_\nu-\partial_\nu A_\mu$. The indices $\mu$ and $\nu$ take the
value 0 and 1 in $(1 + 1)$ dimension. The nature of the
interaction between fermion and gauge field is chiral for this
model. The Jackiw-Rajaram version of bosonized Lagrangian had a
one-parameter class of the covariant mass-like term for the gauge
field that entered into the model in the process of bosonization
in order to remove the singularity in the fermionic determinant
during the course of eliminating the fermion by integration. The
bosonized version of the model (\ref{BCS}) with Faddeevian type of
anomaly \cite{PM, MG, SUBIR1, SUBIR2} is given by
\begin{eqnarray}
{\cal L}_{BCS}&=&\int
dx[\frac{1}{2}\partial_\mu\phi\partial^\mu\phi+
g(\epsilon_{\mu\nu}+g_{\mu\nu})\partial^\nu\phi A^\mu +
\frac{1}{2}g^2(A_0^2 - 2 A_0A_1 -3 A_1^2)
-\frac{1}{4}F_{\mu\nu}F^{\mu\nu}]\nonumber \\
&=&\int dx[ \frac{1}{2}(\dot\phi^2 - \phi'^2) + g(\dot\phi +
\phi')(A_0 - A_1) + \frac{1}{2}g^2(A_0^2 - 2 A_0A_1 -3
A_1^2)+\frac{1}{2}(\dot{A}_1 -A_0')^2]. \label{BCS}
\end{eqnarray}
Here $\epsilon_{\mu\nu}$ is the Levi-Civita  symbol in two
dimension: $\epsilon^{01} = -\epsilon^{10}$=1,  and Minkowski
metric $g_{\mu\nu}= diag(1, -1)$. Equation (\ref{BCS}) was
initially found in \cite{PM} where Mitra termed it as chiral
Schwinger model with Faddeevian regularization since the Gauss law
constraint of this theory gave a specific nontrivial contribution.
We will discuss it later. Let us now discuss the theoretical
spectrum as offered by this model. From the standard definition,
the momentum corresponding to the field $\phi$, $A_1$, and $A_0$
read
\begin{equation}
\pi_\phi = \dot\phi + g(A_0 - A_1),
\end{equation}
\begin{equation}
\pi_1 = \dot{A}_1 - A'_0,
\end{equation}
\begin{equation}
\pi_0\approx 0. \label{MOMA0}
\end{equation}
The following Legendre transformation
\begin{equation}
H_B = \int dx [\pi_\phi\dot\phi + \pi_1\dot{A}_1 - {\cal L}_B],
\end{equation}
leads to the Hamiltonian
\begin{eqnarray}
H_B= \int {\cal H}_B dx = \int dx [\frac{1}{2}(\pi_1^2  +\phi'^2+
\pi_\phi^2) + \pi_1A_0' + g(\pi_\phi + \phi')(A_0 - A_1)+
2g^2A_1^2].
\end{eqnarray}
The gauss law constraint that comes out from the preservation of
$\pi_0\approx 0$ is
\begin{equation}
G = \pi_1' + g(\pi_\phi + \phi')' \approx 0, \label{CONF}
\end{equation}
which has the following nontrivial poisson bracket
\begin{equation} [G(x), G(y)] = 2g^2\delta(x-y)', \label{POIS}
\end{equation}
which we have mentioned already and it is the reason the model is
said to carry the Faddeevian anomaly. Note that this Poisson
bracket (\ref{POIS})gave the vanishing contribution for the
Jackiw-Rajaraman version of the chiral Schwinger model \cite{JR}.
It was the pioneering observation of Faddeev that anomaly made
Poisson bracket between $G(x)$ and $G(y)$ non-vanishing
\cite{FAD1, FAD2}. The constraint became second class itself and
gauge invariance got violated. He, however, argued that it would
be possible to quantize the theory but in this situation, the
system might possess more degrees of freedom. In the article
\cite{PM}, analysis of theoretical spectra with the help of
quantization of constrained system due to Dirac \cite{DIR} was
carried out where it was found that in the phasespce of the model,
along with the constraints (\ref{MOMA0}) and (\ref{CONF}), two
more secondary constraints were embedded in the phasespace, which
were the following:
\begin{equation}
 \pi_1' + 2e\phi' \approx 0,
\end{equation}
\begin{equation}
 A_1' + A_0' \approx 0,
\end{equation}
and these four constraints altogether formed a second class set.
The reduced Hamiltonian of the system obtained after imposition of
the four constraints was found out to be
\begin{eqnarray}
H_r= \int dx {\cal H}_r = \int dx [\frac{1}{2}(\pi_1^2
+\frac{1}{g^2}\pi_1'^2) + \phi'^2 + \pi_1A_1' +
\frac{1}{g}\pi'\phi'+ 2g^2A_1^2].
\end{eqnarray}
The Dirac brackets of the field describing the reduced hamiltonian
were.
\begin{equation}
[\phi(x), \phi(y)]^* = -\frac{1}{4}\epsilon(x-y),
\end{equation}
\begin{equation}
[A(x), A(y)]^*= \delta'(x-y),
\end{equation}
\begin{equation}
[A(x), \phi(y)]^*= \delta(x-y).
\end{equation}
Here $"*"$ indicates the dirac brackets. The theoretical spectrum
contained a massive and a mass less boson chiral boson described
by the following equations
\begin{equation}
[\Box + 4g^2]A_1=0,
\end{equation}
\begin{equation}
(\partial_0 + \partial_1){\cal H}=0, {\cal H}= \phi+\frac
{1}{g}(\dot{A}_1 + A_0').
\end{equation}
Let us now turn to the bosonized vector Schwinger model
\subsection{Vector Schwinger model}
The vector Schwinger model in the fermionic version is defined by
the Lagrangian density
\begin{eqnarray}
{\cal L}_{VS} &=& \bar\psi\gamma^\mu[i\partial_\mu +
eA_\mu]\psi-\frac{1}{4} F_{\mu\nu}F^{\mu\nu}. \label{FVS}
\end{eqnarray}
The bosonized version of the model reads
\begin{eqnarray}
{\cal L}_{BVS} &=& \int
dx[\frac{1}{2}\partial_\mu\phi\partial^\mu\phi+
e\epsilon_{\mu\nu}\partial^{\nu}\phi
A^\mu -\frac{1}{4} F_{\mu\nu}F^{\mu\nu} ],\nonumber \\
&=& \int dx [\frac{1}{2}(\dot\phi^2 - \phi'^2) + e(A_1\dot\phi -
A_0\phi')+ \frac{1}{2}(\dot{A}_1 -A_0')^2].\label{BVS}
\end{eqnarray}
The momenta corresponding to the fields are
\begin{equation}
\pi_{\phi}= \dot\phi + eA_1,\label{MOM1}
\end{equation}
\begin{equation}
\pi_{0}\approx 0, \label{MOM2}
\end{equation}
\begin{equation}
\pi_{1}= \dot{A}_1 -A_0'.\label{MOM3}
\end{equation}
The canonical Hamiltonian is found out to be
\begin{equation}
H_{CV}=\int dx [\frac{1}{2}(\pi_1^2 + \pi_{\phi}^2 + \phi'^2) +
\pi_1A_0' + eA_0\phi'-eA_1\pi_{\phi}]
\end{equation}
The two first class constraints  which are present in the phase
space of the system are
\begin{equation}
\pi_{0}\approx 0, \label{CV}
\end{equation}
\begin{equation}
\pi_1 - e\phi'\approx 0\label{CV1}
\end{equation}
Two gauge fixing conditions chosen here are in order to quantize
the theory such that the real physical degrees of freedom be
identified. The conditions are
\begin{equation}
A_{0} \approx 0, ~~~\pi_{\phi}\approx 0.\label{GFAV}
\end{equation}
The reduced Hamiltonian obtained after imposition of the
constraints and the gauge fixing conditions reads
\begin{equation}
H_R= \int dx [\frac{1}{2}\pi_1^2 + \frac{1}{2e^2}\pi_1'^2
+\frac{1}{2}e^2A_1^2]\label{RHV}
\end{equation}
Here Dirac brackets remain canonical. This reduced Hamiltonian
with the use of canonical Dirac brackets renders the following two
first-order equations of motion:
\begin{equation}
\dot{\pi}_1= -e^2A_1,
\end{equation}
\begin{equation}
\dot{A}_1= \pi_1-\frac{1}{2e^2}\pi''_1.
\end{equation}
The above two equations ultimately reduce to the following two
second order differential equations of motion:
\begin{equation}
[\Box+e^2]\pi_1=0, [\Box+e^2]A_1=0.
\end{equation}
The above two equations  represent a massive boson with mass $e$
and its momentum conjugate.
\subsection{Axia-vector Schwinger model}
The same model with axial vector interaction is described by the
Lagrangian density
\begin{eqnarray}
{\cal L}_{AVS} &=& \bar\psi\gamma^\mu[i\partial_\mu + {\tilde
e}\gamma_5 A_\mu]\psi -\frac{1}{4}
F_{\mu\nu}F^{\mu\nu}.\label{FAVS}
\end{eqnarray}
Here Vector interaction between matter and gauge field is replaced
by axil vector. The bosonized version of the model reads
\begin{eqnarray}
{\cal L}_{BAVS}&=&\int
dx[\frac{1}{2}\partial_\mu\phi\partial^\mu\phi+ {\tilde
e}\partial^\mu\phi A^\mu + \frac{1}{2}{\tilde e}^2 A_\mu
A^\mu-\frac{1}{4}
F_{\mu\nu}F^{\mu\nu}]\nonumber \\
 &=& \int dx[\frac{1}{2}(\dot\phi^2 -
\phi'^2) + {\tilde e} (A_0\dot\phi-
 A_1\phi') ) + \frac{1}{2}{\tilde e}^2(A_0^2 - A_1^2) + \frac{1}{2}(\dot{A}_1 -A_0')^2],
{\label{BAVS}}
\end{eqnarray}
We now proceed to study the phasespace structure of the
axial-vector Schwinger model. To this end, we consider the
bosonized Lagrangian of the axial-vector Schwinger model
(\ref{BAVS}). The momenta corresponding to the fields $\phi$,
$A_0$, and $A_1$ are
\begin{equation}
\pi_{\phi}= \dot\phi + {\tilde e}A_0, \label{AMOM1}
\end{equation}
\begin{equation}
\pi_{0}\approx 0, \label{AMOM2}
\end{equation}
\begin{equation}
\pi_{1}= \dot{A}_1 -A_0'\label{AMOM3}
\end{equation}
A straightforward calculation leads to following  canonical
Hamiltonian.
\begin{equation}
H_{CAV}=\int dx [\frac{1}{2}(\pi_1^2 + \pi_{\phi}^2 + \phi'^2) +
\pi_1A_0' + {\tilde e}A_1\phi'-{\tilde e}A_0\pi_{\phi}].
\end{equation}
Here equation (\ref{AMOM2}) is the primary constraint of the
theory and the preservation of this constraint leads to
\begin{equation}
\pi_1+e\pi_{\phi}\approx 0, \label{CAV}
\end{equation}
which is the secondary constraint. The constraints (\ref{AMOM2})
and (\ref{CAV}) form a first-class set. So like the vector
Schwinger model, two gauge fixing conditions are needed to find
out the physical degrees of freedom. The gauge fixing conditions
are chosen as
\begin{equation}
A_{0} \approx 0, ~~~\phi'\approx 0.\label{GFAVA}
\end{equation}
The reduced Hamiltonian which comes out after imposition of the
constraint and the gauge fixing conditions reads
\begin{equation}
H_R= \int dx [\frac{1}{2}\pi_1^2 + \frac{1}{2{\tilde e}^2}\pi_1'^2
+\frac{1}{2}{\tilde e}^2A_1^2]. \label{RHAV}
\end{equation}
It has been found that the Dirac Brackets are canonical for this
system too. The canonical dirac brackets along with the
Hamiltonian (\ref{RHAV}) lead to the following equations of motion
\begin{equation}
\dot{\pi}_1= -{\tilde e}^2A_1,
\end{equation}
\begin{equation}
\dot{A}_1= \pi_1-\frac{1}{2{\tilde e}^2}\pi''_1.
\end{equation}
The above two equations ultimately reduce to
\begin{equation}
[\Box+{\tilde e}^2]\pi_1=0, [\Box+{\tilde e}^2]A_1=0.
\end{equation}
Like the vector Schwinger model, these two equations also indicate
that the spectrum contains a massive boson with mass $e$ and its
momentum conjugate.

In all the cases the bosonized version of the model is equivalent
to the fermionic version. However in the bosonized version the
models contain a quantum correction because the process of
bosonization involves the integrating out of the fermions from the
actions of the models that lead to fermionic determinant which
carries singularity, and to remove the singularity regularization
is needed. So, the different counter terms may result depending on
the choice of regularization. However, the model with different
counter terms are equivalent to the original fermionic version of
the models
\section{Imposition of chiral constraints on the models}
We have already mentioned that the imposition of chiral constraint
rendered remarkable services in different situations. The
mathematical form chiral constraint, which was introduced in
\cite{KH} was
\begin{equation} \Omega(x) = \pi_\phi(x) -
\phi'(x) \approx 0. \label{CCON}
\end{equation}. It may be of the
form $\Omega(x) = \pi_\phi(x) + \phi'(x) \approx 0$. The
constraint with this form will restrict the degrees of freedom of
the boson with the chirality opposite to the chirality restricted
by the constraint (\ref{CCON}). These are a second-class
constraint since the Poisson bracket of the constraints with
themselves are non-vanishing and the inverse of these exist.
\begin{equation}
[\Omega(x), \Omega(y)] = -2\delta'(x-y). \end{equation} We saw its
remarkable role in the article \cite{KH} where the author was able
to describe the Chiral Schwinger model in terms of Chiral boson.
We are now intended to impose this chiral constraint in the phase
space of the three models defined with three different types of
interaction. The models are discussed in Sec.II. Let us first
consider the chiral Schwinger model with parameter-free Faddeevian
anomaly.
\subsection{Imposition of chiral constraints on the chiral Schwinger
models with parameter-free Faddeevian anomaly}
The bosonized
version of Lagrangian of this model is described in Eqn.
(\ref{BCS}). If we imposing the constraint $\Omega(x) \approx 0$,
into the phasespace of the model the generating functional can be
written down as
\begin{eqnarray}
Z_{CH} = \int d\phi d\pi_\phi  dA_1 d\pi_1  \delta(\pi_\phi -
\phi') \sqrt{det[\Omega(x), \Omega(y)]}e^{ i\int
d^2x[\pi_\phi\dot\phi+ \pi_1 \dot{A}_1 - {\cal H}_B]}
\end{eqnarray}
After a few steps of algebra we land onto to the following:
\begin{eqnarray}
Z_{CH} = N \int d\phi dA_1 e^{i\int d^2x{\cal L}_{CH}},
\end{eqnarray}
Where $N$ stands for the normalization constant, and ${\cal
L}_{CH}$ has the expression
\begin{equation}
{\cal L}_{CH} = \dot\phi\phi' -\phi'^2 + 2g(A_0 - A_1)\phi' + 2g^2
A_1^2 + \frac{1}{2}(\dot{A}_1 -A_0')^2.\label{LCH}
\end{equation}
This can be identified with the gauged Lagrangian for chiral boson
obtained from the bosonized Lagrangian with parameter Faddeevian
regularization \cite{PM} just by imposing the chiral constraint in
its phase space. Harada in \cite{KH}, obtained the same type of
result for the usual chiral Schwinger model with one parameter
class of regularization proposed by Jackiw and Rajaraman
\cite{JR}. The Lagrangian (\ref{LCH}) can be thought of as the
gauged version of chiral boson described by Floreanini and Jackiw
\cite{FLO}.
\subsection{Imposition of chiral constraints on the vector Schwinger
models}
 Let us now impose the chiral constraint in the pasespace
of the vector Schwinger model. The bosonized version of the
Lagrangian of this model is given in Eqn. (\ref{BVS}). The chiral
constraint which was imposed in the chiral Schwinger model with
the Faddeevian anomaly was
\begin{equation}
 \Omega(x) = \pi_\phi(x) -
\phi'(x) \approx 0.
\end{equation}
The Poisson bracket of this  constraint with itself is
non-vanishing as usual
\begin{equation}
[\Omega(x), \Omega(y)] = -2\delta'(x-y).
 \end{equation}.
So it is a second class constraint. The way Harada imposed the
chiral constraint \cite{KH} in the chiral Schwinger model can be
extended to the Schwinger model without any hindrance since no
physical principle will be violated here too like the Chiral
Schwinger model: All though the nature of the interaction is
different in the models, in the kinetic term boson's of both the
chirality are present and the constrain in Eqn. (\ref{CCON}) puts
restriction on the kinematics of the boson. If we imposing the
constraint $\Omega(x) \approx 0$, the generating functional
corresponding this model can be written down as follows:
\begin{eqnarray}
Z_{CHV} = \int d\phi d\pi_\phi dA_1 d\pi_1 \delta(\pi_\phi -
\phi') \sqrt{det[\Omega(x), \Omega(y)]}e^{ i\int
d^2x[\pi_\phi\dot\phi+\pi_1\dot{A}_1 - {\cal
H}_{BVS}]}\end{eqnarray} A straightforward calculation leads to
\begin{eqnarray}
Z_{CHV} ={\tilde N}\int d\phi d A_1e^{i\int d^2x{\cal L}_{CHV}},
\end{eqnarray}
where ${\tilde N}$ is the normalization constant. The Lagrangian
density ${\cal L}_{CHV}$ reads
\begin{equation}
{\cal L}_{CHV} = \dot\phi\phi' -\phi'^2 + 2e(A_0 - A_1)\phi' +
2e^2 A_1^2 + \frac{1}{2}(\dot{A}_1 -A_0')^2. \label{LCHV}
\end{equation}
It is gauge non-invariant Lagrangian here Lorentz co-variance is
not manifested.
\subsection{Imposition of chiral constraints on the axial vector Schwinger
models}
 The bosonized version of the Lagrangian of the
axial-vector Schwinger model is given in  Eqn. (\ref{BAVS}). Let
us now impose the same chiral constraint $\Omega(x) = \pi_\phi(x)
- \phi'(x) \approx 0$ in the phasespace of the axial-vector
Schwinger models. If we do so the generating functional of the
axial-vector Schwinger will be written down as
\begin{eqnarray}
Z_{CHA}=\int d\phi d\pi_\phi  dA_1 d\pi_1 \delta(\pi_\phi - \phi')
\sqrt{det[\Omega(x), \Omega(y)]} e^{ i\int d^2x[\pi_\phi\dot\phi+
\pi_1 \dot{A}_1 - {\cal H}_{BACS}]}.
\end{eqnarray}
After a little algebra it reduces to
\begin{eqnarray}Z_{CHA} =\bar{N}\int
d\phi d A_1 e^{i\int d^2x{\cal L}_{CHA}},
\end{eqnarray}
Here ${\cal L}_{CHA}$ is given by
\begin{equation}
{\cal L}_{CHA} = \dot\phi\phi' -\phi'^2 + 2{\tilde e}(A_0 -
A_1)\phi' + 2{\tilde e}^2 A_1^2+ \frac{1}{2}(\dot{A}_1 -A_0')^2,
\label{LCHA}
\end{equation}
and $\bar{N}$ represents the normalization constant. It is also a
gauge non-invariant Lagrangian where Lorentz co-variance is not
manifested. It is important to note that the Lagrangians obtained
after imposition of chiral constraint in the phasespaces of the
Chiral Schwinger model with parameter-free Faddeevian anomaly,
Vector Schwinger model, and the axial-vector Schwinger model are
not different. It is surprising indeed that three models with
different interactions become indistinguishable after the
imposition of chiral constraint. It is undoubtedly a novel aspect
of imposition of chiral constraint
\section{Description of theoretical spectra of the resulting model
obtained after imposition of the chiral constraint}
The constraint structure and the theoretical spectra of this model
is known \cite{PM}. The primary constraint of the theory are
\begin{equation}
\Omega_1 =\pi_0\approx 0
\end{equation}
\begin{equation}
\Omega_2= \pi_\phi-\phi'\approx 0
\end{equation}
The canonical Hamiltonian is
\begin{eqnarray}
H_{CB}= \int {\cal H}_{CB} dx = \int dx [\frac{1}{2}\pi_1^2
+\phi'^2 + \pi_1A_0' - 2g\phi'(A_0 - A_1)+ 2g^2A_1^2].
\end{eqnarray}
The secondary constraints are
\begin{equation}
\Omega_3 = {\pi}'_{1} + 2g{\phi}'\approx 0, \label{AKIK30}
\end{equation}
\begin{equation}
\Omega_4 = (A_1 + A_0)' \approx 0.\label{AKIK40}
\end{equation}
The reduced Hamiltonian reads
\begin{eqnarray}
H_{CR}= \int dx {\cal H}_{CR} = \int dx [\frac{1}{2}\pi_1^2
+\frac{1}{4g^2}\pi_1'^2 + \pi_1A_1' +  2g^2A_1^2].
\end{eqnarray}
The non-canonical bracket reads
\begin{equation}
[A_1(x), A_1(y)]=\frac{1}{2g^2}\delta'(x-y) = 0
\end{equation}
The theoretical spectra contains only a massive boson with mass
 $2g$
\begin{equation}
[\Box + 4g^2]A_1 = 0
\end{equation}
\section{Gauge invariant reformulation using Mtra-Rajaraman's formalism}
Mitra and Rajaraman developed an ingenious formalism in their
seminal work \cite{MR1, MR2} for obtaining a gauge-invariant
theory by reducing the number of constraints from a second class
set of constraints belonging to a theory retaining only the
first-class subset. The remarkable feature of this formalism is
that the gauge invariance is received in the usual phasespace of
the theory. Unlike Stueckelberg formalism \cite{STU1, STU2, STU3},
the extension of phasespce is not required here. For a gauge
theory fixing of the gauge is required in order to single out the
physical degrees of freedom and it is done by imposing a suitable
number of gauge fixing conditions. The first-class set of
constraints that the theory is endowed with form a second class
set together with the gauge fixing condition. As a result, the
theory turns into an equivalent second-class system. In the
article \cite{MR1, MR2}, an inverse to the above procedure is
invoked that enables to have a first-class gauge-invariant system
corresponding to a second class theory gauge variant theory.

The reduction of the number of constraints is done in such a way
that the constraints which are eliminated may be thought of as the
gauge fixings of the first-class set of constraints that retains
in. This formulation crucially depends on the constraints that
embed in the phasespace of the theory and different
gauge-invariant version may result which solely depend on which
set of first-class constraints are retained. Since no extension of
phase space is done here the physical contents of the resulting
gauge-invariant actions remain unaltered. What follows next is the
use of this formalism to have a gauge-invariant version of the
chiral Schwinger model with parameter-free Faddeevian
regularization when it is described in terms of chiral boson
\cite{PM}

This model described in (\ref{LCH}) contain two primary
constraints
\begin{equation}
\Omega_1 = \pi_{0}\approx 0,\label{AKIK1}
\end{equation}
\begin{equation}
\Omega_2=\pi_\phi-\phi' \approx 0. \label{AKIK2}
\end{equation}.
The effective Hamiltonian of this system is therefore given by
\begin{equation}
H_{eff}= H_{CB} + v_1 \pi_0 + v_2(\pi_\phi-\phi'),
\end{equation}
where the canonical Hamiltonian $H_{CB}$ is given by
\begin{equation}
H_{CB}= \int dx[\frac{1}{2}\pi_1^2 + \pi_1 A_0' + \phi'^2 -2g(A_0
-A_1)+2g^2A_1^2]
\end{equation}
There are two secondary constraints  in the phase space of the
theory which are the following
\begin{equation}
\Omega_3 = {\pi}'_{1} + 2g{\phi}'\approx 0, \label{AKIK3}
\end{equation}
\begin{equation}
\Omega_4 = A_1' + A_0' \approx 0.\label{AKIK4}
\end{equation}
The Poisson bracket between  $\Omega_1\approx 0$ and $\Omega_3
\approx 0$ vanishes. Therefore, these two constraints form a
first-class subset from the full set of four constraints.
Following the formalism developed in \cite{MR1, MR2}, if we are
intended to retain only these two constraints (\ref{AKIK1}), and
(\ref{AKIK2}) the Hamiltonian needs the following modification
\begin{eqnarray}
H_{MOD}&=&\int dx[
\frac{1}{2}\pi_1^2+\pi_1A_0'-2e(A_0-A_1)\phi'+2g^2A_1^2+
\pi_{\phi}\phi'-e\pi_{\phi}(A_0-A_1)  \nonumber \\
 &+&  \frac{1}{2}(\pi_{\phi}-\phi')^2 +g(\pi_{\phi}-\phi')(A_0 + A_1)
+ w\pi_0].\label{MODHA}
\end{eqnarray}
Note that this modified Hamiltonian (\ref{MODHA}) retains only two
first-class constraints $\Omega_1 \approx 0$ and $\Omega_3 \approx
0$, because the preservation of the constraint $\Omega_1 \approx
0$, and $\Omega_3 \approx 0$ with respect to the modified
Hamiltonian does not lead to any new constraint and it does not
alter the physical contents of the theory since the modified
Hamiltonian contains only those fields with which it was defined.
The equations of motion that follow from the modified first-class
Hamiltonian(\ref{MODHA}) are
\begin{equation}
\dot{\phi} = [\phi, H_{MOD}] = \pi_{\phi}+ 2gA_1 ,\label{EQN1}
\end{equation}
\begin{equation}
\dot{A_0}= [A_0, H_{MOD}]= -u\label{EQN2},
\end{equation}
\begin{equation}
\dot{A_1} = [A_1, H_{MOD}] = \pi_1\label{EQN3}.
\end{equation}
Through a Legendre transformation we obtain the Lagrangian
corresponding to the modified Hamiltonian (\ref{MODHA})
\begin{eqnarray}
L &=&\int dx[
\pi_{\phi}\dot{\phi}+\pi_{1}\dot{A_1}+\pi_{0}\dot{A_0}-(\frac{1}{2}\pi_1^2
+\pi_1A_0' +2gA_1\pi_{\phi} \nonumber \\
&+&\pi_{\phi}\phi'-2gA_0\phi'
+\frac{1}{2}(\pi_{\phi}-\phi')^2+2g^2A_1^2+w\pi_0)]. \label{LAGF}
\end{eqnarray}
By the use of equations (\ref{EQN1}), (\ref{EQN2}), and
(\ref{EQN3}) Lagrangian (\ref{LAGF}) is expressed in a simplified
form after a little algebra:
\begin{eqnarray}
L&=&\int dx[\frac{1}{2}(\dot{\phi}^2-\phi'^2) -2g(A_1\dot{\phi}
-A_0\phi')+\frac{1}{2}(\dot{A_1}-A_0')^2]
\nonumber \\
&=& \int dx[\frac{1}{2}\partial_\mu\phi\partial^\mu\phi+
2g\epsilon_{\mu\nu}\partial^\nu\phi A^\mu -\frac{1}{4}
F_{\mu\nu}F^{\mu\nu}] \label{FLAG}
\end{eqnarray}
The Lagrangian(\ref{FLAG}), corresponds to the
Hamiltonian(\ref{MODHA}), and it has the identical set of
equations of motion (\ref{EQN1}), (\ref{EQN2}), and (\ref{EQN3}).
It is straightforward to see that the Lagrangian (\ref{FLAG}) has
two primary constraints (\ref{AKIK1}) and (\ref{AKIK2} only. This
Lagrangian can easily be identified with the bosonized version of
the well-celebrated vector Schwinger model \cite{SCH, LOW}. Here
coupling strength is $2g$. We can set $g=\frac{1}{2}e$ to make it
identical to with equation (\ref{BVS}).

It is known from the seminal work of Dirac \cite{DIR} that the
gauge transformation generator is constructed from the first-class
constraints of a theory. The gauge transformation generator in
this situation can be written down as
\begin{equation}
{\cal G} = \int dx(\Lambda_1\Omega_1 + \Lambda_2\Omega_3).
\label{GENERATOR}
\end{equation}
Here $\Lambda_1$ and $\Lambda_2$ are two arbitrary parameters that
will be fixed later. The gauge transformation for the fields
$\phi$, $A_1$, and $A_0$ that stems out from the generator read
\begin{equation}
\delta \phi =0, \delta A_1 = -\Lambda_1',  \delta A_0 =
-\Lambda_2. \label{TRANSF}
\end{equation}
It is straightforward to see that the under the transformation
(\ref{TRANSF}), the Lagrangian (\ref{FLAG}) remains invariant if
the the parameters $\Lambda_1$ and $\Lambda_2$ are constrained to
\begin{equation}
\Lambda_2=\dot{\Lambda_1}.
\end{equation}
This transformation is equivalent to the familiar form of the
gauge transformation $A_\mu \rightarrow A_\mu +
\frac{1}{2g}\partial_\mu\Lambda$.
There is something interesting
that we must mention here.  Note that the gauge-invariant version
of the chiral Schwinger model with parameter-free Faddeevian
regularization land on to the celebrated Schwinger model and the
Schwinger model under a chiral constraint lands onto  the chiral
Schwinger model with the parameter-free Faddeevian regularization.
It is surprising that the nature of interaction of the two models
to start with was different. However the imposition of the chiral
constraint in the phase-space makes the nature of interaction
indistinguishable and that ultimately lead identical theoretical
spectra.
\section{Discussion and Conclusions}
In this article we consider three different models having
different types of interaction between the matter and the gauge
field. The models are with parameter-free Faddeevian anomaly, the
vector Schwinger model, and the axial-vector Schwinger model. We
should mention here that the Schwinger itself can be described
with two different types of interaction namely vector and
axial-vector interaction. We have used the bosonized version of
all these models, which contain quantum corrections that enter
through the regularization needed in order to remove the
singularity in the fermionic determinant that appears during the
process of integrating out the fermions. So different
counter-terms result because of the choice of regularization. For
the vector and axial-vector Schwinger model we consider the usual
regularization. However, for the Chiral Schwinger model, we use
the parameter-free Faddeevian regularization developed in
\cite{PM, MG, SUBIR1, SUBIR2}. We impose chiral constraint
following the article \cite{KH} in the phase space of all these
three models and found that all the three models map onto a gauge
non-invariant model and it has no manifestly Lorentz covariant
structure, but it is known from the article \cite{MG}, that the
model exhibits Lorentz invariant spectrum. It is also observed
that the gauge-invariant version of the resulting model maps onto
the vector bosonized version of the vector Schwinger model.

It is surprising that three different models having different
types of interaction between the matter and gauge field become
indistinguishable under a chiral constraint at the quantum
mechanical level. To unveil the precise reason for obtaining this
the remarkable results need further involved investigations.
However, some comments can be made. The models considered here are
described in (1+1) dimensional spacetime manifold. The constraint
structure and the constrained subspace of the models are such that
under this specific chiral constraint the physical phasespace
become indistinguishable and that, in turn, result in the
identical theoretical spectra.

Besides, the imposition of chiral constraint in the vector
Schwinger model, axial-vector Schwinger model, and chiral
Schwinger model with parameter-free Faddeevian anomaly leads to
single gauge non-invariant model which has a gauge invariant
structure in the same phase space that can be identified with the
gauge invariant vector Schwinger model. It also favors the
indistinguishable nature of the models.

\end{document}